# Dual-Frequency Resonance-Tracking Atomic Force Microscopy

Brian J. Rodriguez,[1,2] Clint Callahan,[3] Sergei V. Kalinin,[1,2,*] and Roger Proksch[3,†]

[1]Materials Science and Technology Division and [2]The Center for Nanophase Materials Sciences, Oak Ridge National Laboratory, Oak Ridge, TN 37831

[3]Asylum Research, Santa Barbara, CA 93117

A dual-excitation method for resonant-frequency tracking in scanning probe microscopy based on amplitude detection is developed. This method allows the cantilever to be operated at or near resonance for techniques where standard phase locked loops are not possible. This includes techniques with non-acoustic driving where the phase of the driving force is frequency and/or position dependent. An example of the later is Piezoresponse Force Microscopy (PFM), where the resonant frequency of the cantilever is strongly dependent on the contact stiffness of the tip-surface junction and the local mechanical properties, but the spatial variability of the drive phase rules out the use of a phase locked loop. Combined with high-voltage switching and imaging, dual-frequency, resonance-tracking PFM allows reliable studies of electromechanical and elastic properties and polarization dynamics in a broad range of inorganic and biological systems, and is illustrated using lead zirconate-titanate, rat tail collagen, and native and switched ferroelectric domains in lithium niobate.

[*] Corresponding author, sergei2@ornl.gov
[†] Corresponding author, roger@asylumresearch.com

In scanning probe microscopy (SPM), changes in the resonance frequencies of the cantilever as it interacts with the surface can provide a direct measure of tip-surface interactions. Following the seminal work by Albrecht et al,[1] resonant frequency detection has emerged as the functional basis of techniques such as non-contact Atomic Force Microscopy (NC-AFM)[2] and contact techniques like Atomic Force Acoustic Microscopy (AFAM).[3] Frequency tracking in these examples is typically implemented using Phase-Locked Loop (PLL) circuitry or logic that utilizes the measured phase lag between excitation and response signals as the error signal for a feedback loop that maintains the cantilever phase at a constant value (typically 90 degrees at resonance) by adjusting the frequency of the excitation signal. The high sensitivity of PLL-based frequency tracking methods allows precise control of tip-surface interactions, enabling atomic-resolution imaging and dissipation probing in NC-AFM. Furthermore, resonance enhancement minimizes contributions of non-thermomechanical noise sources (e.g. $1/f$ and laser shot noise) to the signal, increasing the signal to noise ratio and allowing weak conservative and dissipative tip-surface interactions to be probed.

The PLL methods are ultimately based on the fact that for a constant driving force, the phase and amplitude of the system response are directly linked to proximity to the resonance. The inherent limitations of phase-detection based resonant frequency tracking methods are that the phase of the excitation signal should be (i) independent of the tip position and (ii) independent of the drive frequency. Any measurement where one or both of these conditions is not met will at least limit the interpretability if not the stability of a PLL.

The first condition is typically met for techniques based on acoustic excitation of the cantilever base (NC-AFM) or the sample (AFAM). Depending on the drive mechanism however, the second condition may not be met and the phase of the drive may vary significantly as the resonance frequency changes. In methods based on the electrical excitation of the tip such as Kelvin Probe Force Microscopy (KPFM)[4] and Piezoresponse Force Microscopy (PFM),[5] the relationship between the phase of the excitation force and driving voltage strongly depends on material properties, violating condition (i) above. For example, in PFM the phase of the cantilever response changes by 180º across antiparallel domain walls, and thus cannot be reliably used as a feedback signal, as illustrated in Fig. 1.

As with many SPM techniques, PFM can benefit from an enhanced signal to noise level if the cantilever is operated on resonance.[6,7] However, since PFM also relies on the relative phase of the drive and response for determining domain alignment, there can be significant crosstalk between the sample topography (and by extension, the contact stiffness) and the measured phase of the cantilever. This effect is especially problematic when the cantilever is operated at or near a resonance.[8,9] Previously, a resonance-enhanced PFM based on the rapid detection of an amplitude-frequency response curve at each pixel (10-100 ms per pixel) has demonstrated the need for independent detection of resonant frequency and electromechanical response.[10] Here, we develop and implement a dual-frequency excitation method[11,12] that allows amplitude-based resonant frequency tracking in PFM and similar techniques.

The principle of the method is illustrated schematically in Fig. 2 (a). In this apparatus, the potential of the conductive cantilever is oscillated at a sum of two voltages with frequencies at or near the resonance. The resulting cantilever deflection is digitized and then sent to two separate lock-in amplifiers, each referenced to one of the drive signals. By measuring the amplitudes at these two frequencies, it is possible to measure changes in the resonance and furthermore, to track the resonant frequency. The amplitude-frequency curve for a typical surface tune is schematically indicated by the solid line in Figure 2(b). In the Dual-Frequency Resonance-Tracking (DFRT) method described herein, this response is measured at two drive frequencies, $f_1$ and $f_2$, yielding amplitudes $A_1$ and $A_2$. The difference between these two frequencies $\Delta f = f_2 - f_1$ is typically chosen such that $\Delta f \geq 2BW$ where $BW$ is the imaging bandwidth, typically on the order of 1 kHz. For the cantilevers used in this work, it typically implied amplitudes $A_1, A_2 \sim A_r/2$, where $A_r$ is amplitude at the resonance. The drive frequencies were also typically chosen such that $A_2 - A_1 \approx 0$, though this is not a requirement of the technique.

Referring again to Fig. 2 (b), a change in the contact stiffness of the tip-surface contact during scanning results in a shifted response curve shown as a dashed line. When the response curve changes, the measured amplitudes become $A_1'$ and $A_2'$ respectively. The decrease (increase) of the resonant frequency results in the decrease (increase) of the amplitude difference signal, $A_2' - A_1' < 0$ ($A_2' - A_1' > 0$). Hence, the amplitude difference signal can be used as an input to a feedback loop to maintain the two drive frequencies bracketing the resonant frequency of the cantilever. In our implementation of this method, the two frequencies $f_1$ and $f_2$, chosen such that $\Delta f$ is constant, are updated to maintain $A_2 - A_1 = 0$. We accomplish this using a digital proportional-integral gain controller, although other feedback loop implementations are certainly possible. For a symmetric peak, the resonant frequency is then determined as $f_c = (f_2 + f_1)/2$. For a constant driving force (e.g., acoustic excitation), it is also possible to quantify damping and non-conservative tip-surface interactions from the measured amplitude(s). For PFM however, determination of the damping requires a measurement at an additional frequency (for example, the peak amplitude), to be reported elsewhere.[13] Note that many more forms of frequency feedback based on multiple excitation signals can also be implemented. Examples include more complex functions of the measured amplitudes, phases, in-phases and quadrature components, the cantilever deflection, lateral and/or torsional motion.

We implemented DFRT-PFM on a commercial SPM system (Asylum Research MFP-3D). Measurements were performed using Pt-Ir coated (Olympus Electrilevers) and Au coated (Olympus TR400PB) cantilevers. The system is additionally equipped with a prototype high voltage module (power supply, amplifier, tip holder, and sample holder) that allows application of dc voltages up to +/-220 V and imaging at ac voltages up to 110 $V_{pp}$ (in the dual excitation mode) at frequencies of 300-400 kHz. This enables polarization switching in high-coercitivity materials as single-crystal periodically poled $LiNbO_3$ and imaging of weakly-piezoelectric material (~1-5 pm/V) such as dentin, collagen, and other biopolymers.

Figure 3 illustrates DFRT-PFM imaging of a model lead zirconate-titanate (PZT) polycrystalline surface. The deflection image in Fig. 3 (a) shows several topographic steps clearly visible in differential contrast. The corresponding resonant frequency image exhibits significant (~2 kHz) variations of the contact resonant frequency associated with topographic variations due to changes in the tip-surface spring constant. Large-scale variations in the frequency image, presumably due to changes in elasticity from grain to grain in the bulk ceramic sample and/or surface contamination are also visible on larger images (not shown). Note that for resonant frequencies of ~310 kHz and typical $Q$ factors on the order of ~100, the width of the resonant peak is ~ 3 kHz. This has two implications we mention here; (i) operating near resonance yields a ~10-100-fold increase in the electromechanical response over non-resonant, constant-frequency PFM while (ii) at the same time avoiding crosstalk between changes in the contact stiffness and the PFM signal.

Because we take advantage of the natural cantilever resonance and avoid artifacts associated with changes in the contact stiffness, DFRT-PFM is a promising technique for high-resolution imaging of weakly piezoelectric biopolymers in e.g., calcified and connective tissues.[14,15] While the combination of optical activity and polar bonding renders piezoelectricity ubiquitous in these materials, high-veracity structural imaging by PFM was demonstrated to date only for dentin and artificially prepared collagen films.[15] For most biosystems, surface topographic features strongly correlate with the molecular orientation, precluding unambiguous separation of intrinsic electromechanical response and topographic effects. DFRT-PFM imaging of rat tail collagen is illustrated in Fig. 4. The deflection image in Fig. 4 (a) clearly shows the characteristic periodicity of collagen fibrils. The corresponding resonant frequency map in Fig. 4 (b) shows strong variations of the resonant frequency along the fibril, related to the variation of local stiffness of the sample. The grooves on the fibril are associated with a significant increase of the resonant frequency (~3-4 kHz), consistent with an increase of local contact area and contact stiffening. In addition, strong variations in resonant frequency are visible perpendicular to the fibril axis. Circled in Fig. 4(b) is a region with significantly depressed (~4-5 kHz) local resonant frequency, corresponding to a suspended segment of the collagen fibril. This softening is observed despite the fact that the fibril radius (~150 nm) is significantly larger than the tip radius (~10-20 nm from resolution). This observation strongly suggests that the elastic response of the fibril is non-local, indicative of a softer internal core surrounded by stiffer outer shell. Finally, the PFM images in Fig. 4 (c,d) illustrate only weak amplitude contrast and relatively small (~20°) phase variations within the fibril. This behavior is consistent with that expected for a non-piezoelectric material. The response in this case is the surface deformation induced by the tip-surface electrostatic forces, rather than intrinsic electromechanical coupling in the material, and frequency data can be analyzed similar to AFAM.[3] The lack of measurable electromechanical response in the collagen fibril is not surprising and can be ascribed to the surface being covered by a proteoglycan layer that precludes electrical tip-surface contact, and possibly surface conductivity of humid collagen.

Finally, DFRT-PFM studies of polarization switching are illustrated for single-crystal lithium niobate (LN) (Crystal Technologies). The resonant frequency image of the LN surface in Fig. 5 (a) illustrates a slight frequency shift between domains. Corresponding PFM amplitude and phase images in Fig. 5 (b,c) showing macroscopic 180° domain wall and two inversion domains are typical for this material. Higher resolution DFRT-PFM images of preexisting domains in Fig. 5 (d-f) illustrate strong frequency contrast, and nearly constant PFM amplitudes within and outside the domain. In comparison, shown in Figs. 5 (g-i) are DFRT-PFM images of domains switched by the application of three 176 V magnitude pulses for ~ 10 s in three adjacent locations. Note the significant change of resonant frequency and the strong amplitude depression in the newly fabricated domain. This contrast can be attributed to the transient behavior of the LN surface due to defect-dipole reorientation[16] and surface screening[17] strongly modifying electromechanical response. These observations suggest that the temporal evolution of resonant frequencies and electromechanical responses in systems with screening (all ferroelectrics in ambient), or phase separation (relaxor ferroelectrics) dynamics can provide new approaches to study these phenomena.

To summarize, we have developed a dual-excitation method for resonant-frequency tracking in SPM based on amplitude detection. This approach allows the limitations of frequency tracking in techniques with strongly position-dependent response phase to be overcome. This approach is implemented for PFM, and imaging at 0.3 - 1 Hz rates with < 0.2 kHz frequency noise is demonstrated. These characteristics make this approach ideal for techniques such as PFM, AFAM, and polarization and dielectrophoretic force microscopies in liquid, in which resonant frequencies change significantly (~0.1-40 kHz) across the surface. DFRT-PFM imaging of model PZT surface showed strong variation of resonant frequency at the topographic steps and lack of topographic effects in resonant PFM amplitude, illustrating effective decoupling of the two. DFRT-PFM of collagen fibrils have shown clear elastic contrast within the fibril at 64 nm periodicity, and between suspended and bound segments. At the same time, the fibril is non-piezoelectric within the detection limit, presumably due to the surface proteoglycan layer or surface conductivity. Finally, the strong effects of surface chemistry and screening conditions on the resonant frequency of the PFM cantilever on a LN surface is shown, paving the path way for dynamic studies of slow polarization-mediated phenomena. Future development of this method will include detection at more than 2 frequencies, necessary for unambiguous determination of damping in PFM-type experiments and allowing measurements of tip-sample damping variations.

Research supported in part (BJR, SVK) by the Division of Materials Science and Engineering, Oak Ridge National Laboratory, managed by UT-Battelle, LLC, for the U.S. Department of Energy. BJR acknowledges the financial support of Asylum Research and the use of Asylum Research imaging facilities.

# Figure Captions

**Fig. 1.** Schematic showing a phase shift for antiparallel domains demonstrating that phase cannot reliably be used as a feedback signal.

**Fig. 2.** (a) Schematics of the experimental set-up. (b) Principle of the dual-frequency excitation based resonant-amplitude tracking.

**Fig. 3.** (a) Topography error signal, (b) resonance frequency, (c) piezoresponse amplitude and (d) piezoresponse phase images of the lead zirconate-titanate (PZT) ceramic surface. The images are obtained at $\Delta f = 6.5$ kHz and $V_{ac} = 44$ V.

**Fig. 4.** (a) Topography error signal, (b) resonance frequency, (c) piezoresponse amplitude and (d) piezoresponse phase images of the mouse tail collagen. The images are obtained using high-voltage PFM module at $\Delta f = 10$ kHz and $V_{ac} = 66$ V.

**Fig. 5.** (a,d,g) Resonance frequency, (b,e,h) piezoresponse amplitude and (c,f,i) piezoresponse phase images of the periodically poled lithium niobate surface. Shown are images of the (a-c) native domain structure, (d-f) an intrinsic domain, and (g-i) domains switched by +/- 176 V (locations marked in (e)). The images are obtained at $\Delta f = 4$ kHz and $V_{ac} = 66$ V. The frequency images have been flattened to account for minute changes of contact radius from line to line.

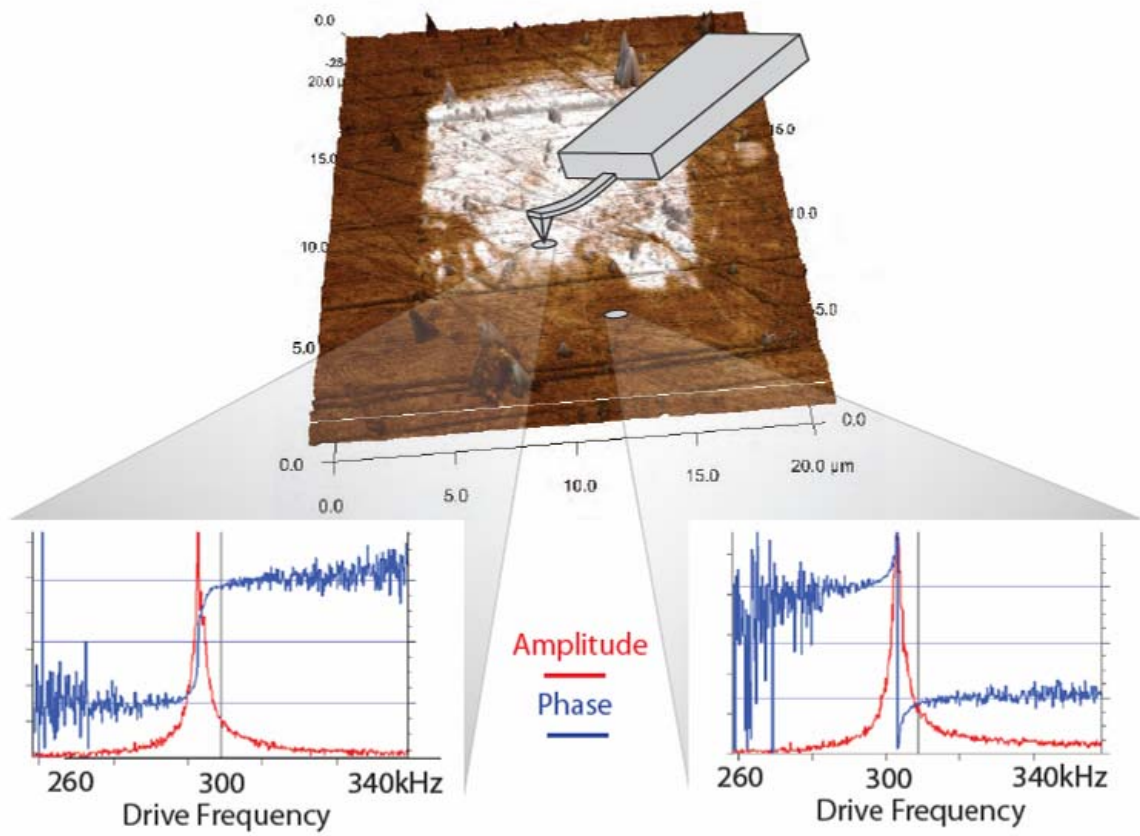

**Fig. 1.** B.J. Rodriguez *et al.*

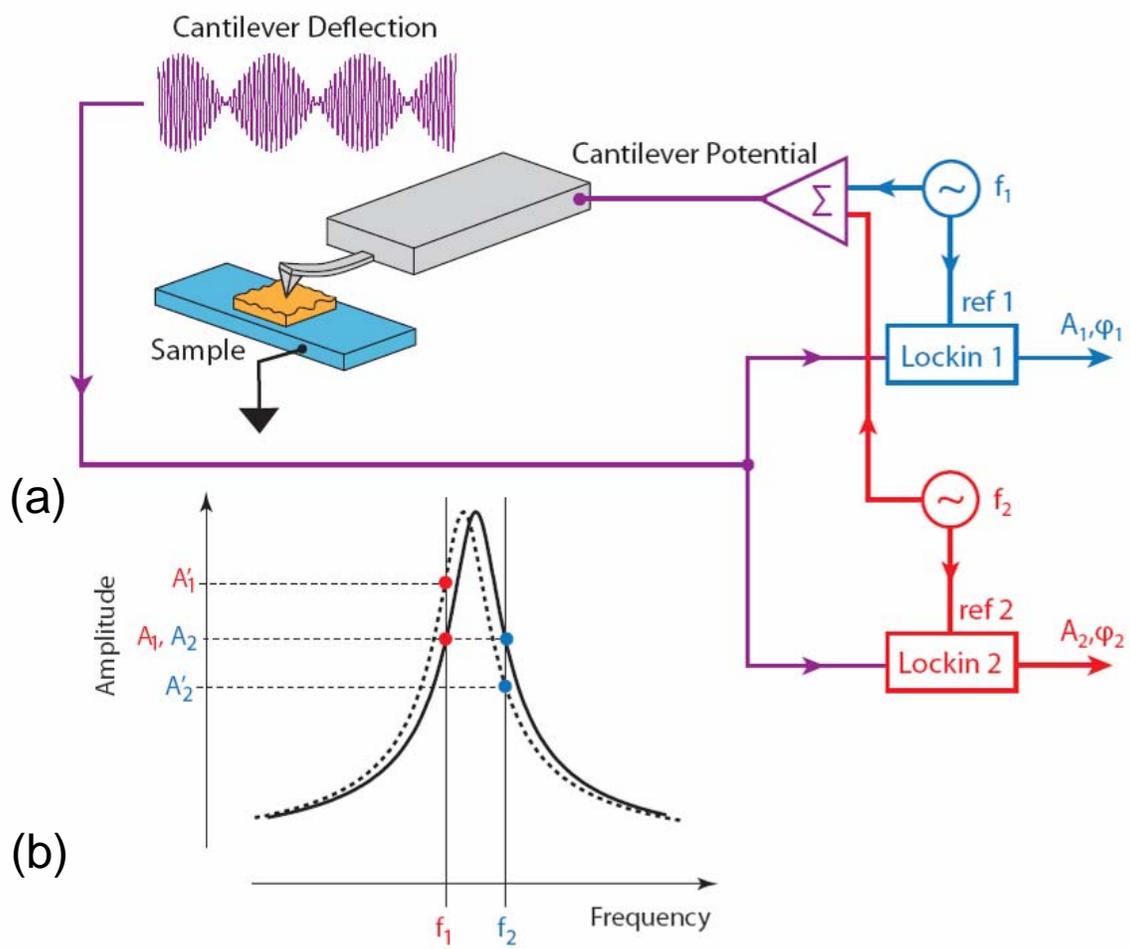

(a)

(b)

**Fig. 2.** B.J. Rodriguez, *et al.*

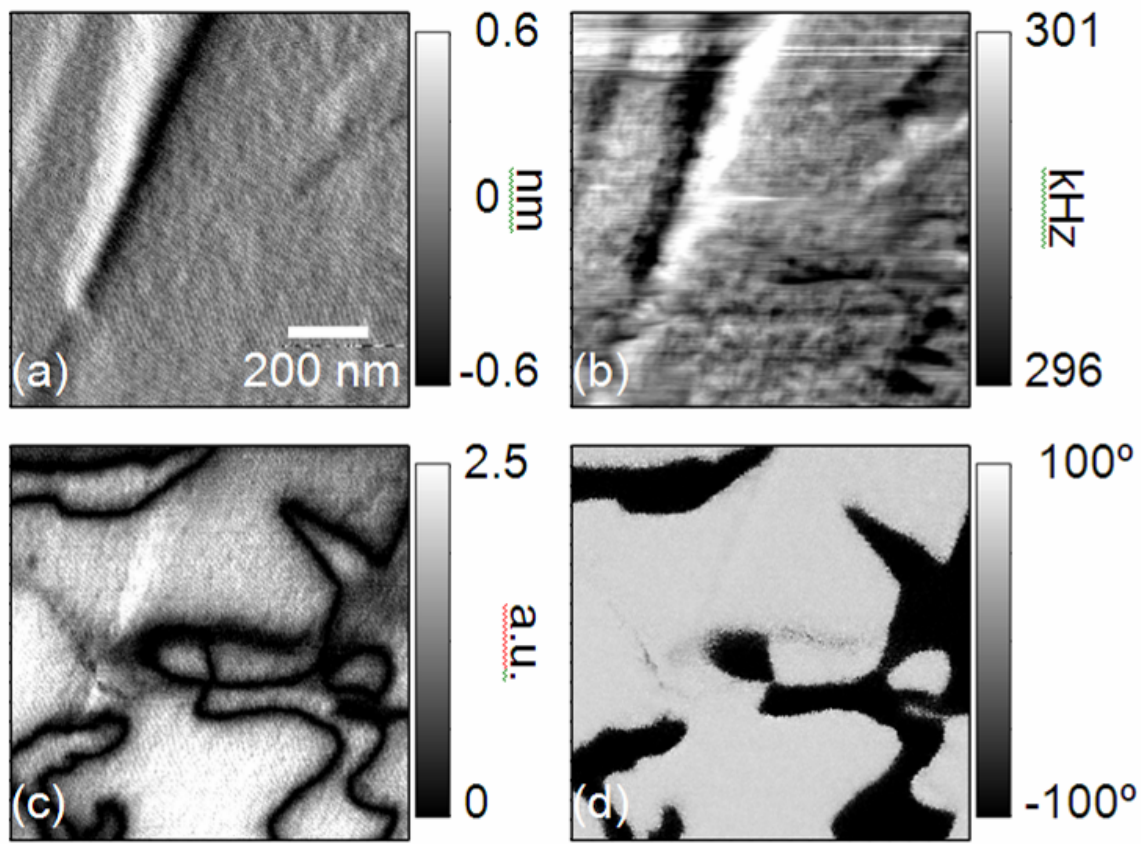

**Fig. 3.** B.J. Rodriguez, *et al.*

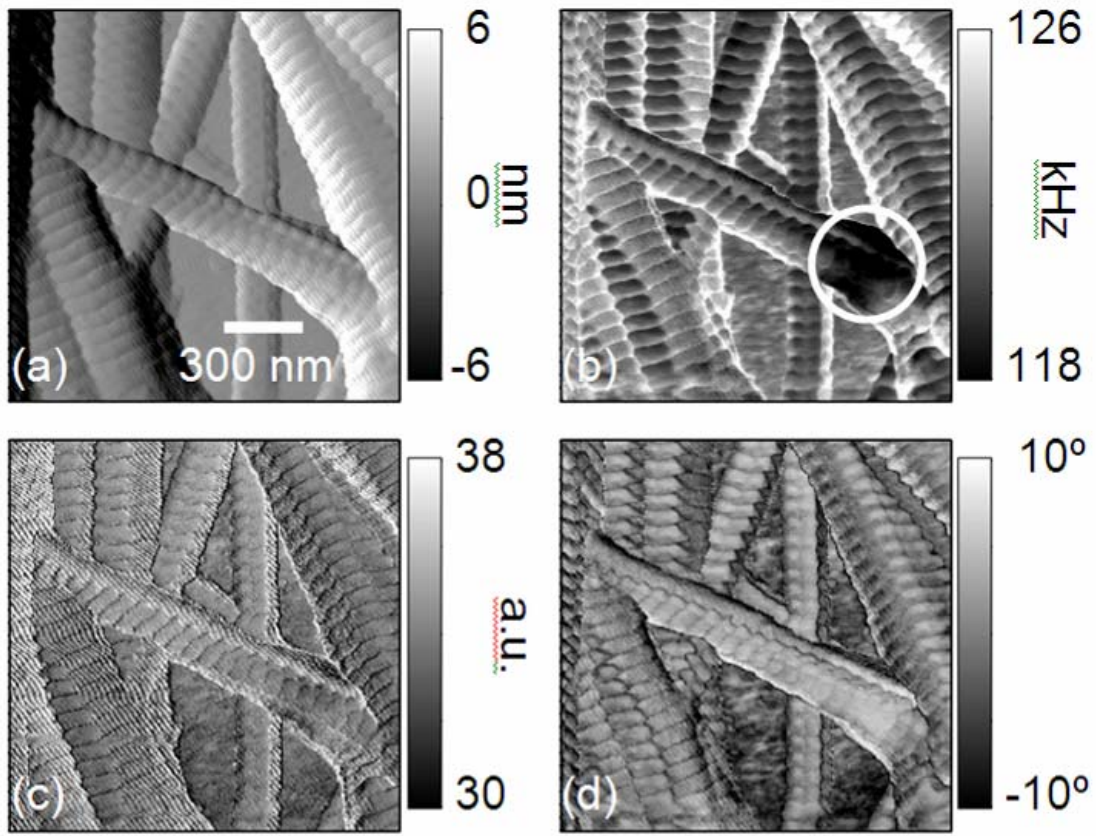

**Fig. 4.** B.J. Rodriguez *et al.*

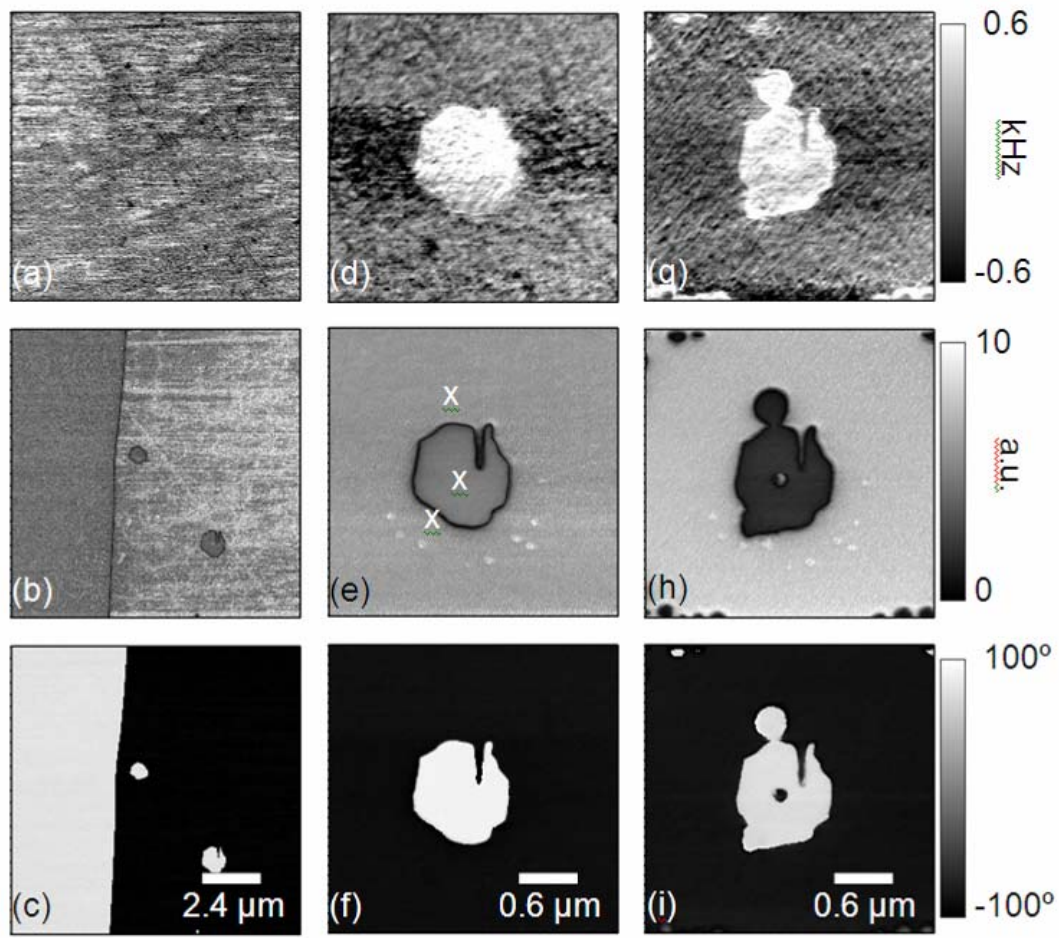

**Fig. 5.** B.J. Rodriguez *et al.*